\documentclass[aps,pre,showpacs,superscriptaddress,twocolumn]{revtex4-1}

\usepackage{graphicx}
\usepackage{amsmath}
\usepackage{amssymb}
\usepackage{bm}

\begin{document}

\title{Origin of the Growing Length Scale in  $M$-$p$-Spin Glass Models }

\author{Joonhyun Yeo}\email{jhyeo@konkuk.ac.kr}
\affiliation{Division of Quantum Phases and Devices,
School of Physics, Konkuk University, Seoul 143-701, Korea}
\affiliation{School of Physics, Korea Institute for Advanced Study, Seoul 130-722, Korea}
\author{M. A. Moore}\email{m.a.moore@manchester.ac.uk}
\affiliation{School of Physics and Astronomy, University of Manchester,
Manchester M13 9PL, UK}
\date{\today}

\begin{abstract}
Two versions  of the $M$-$p$-spin  glass model have been  studied with
the  Migdal-Kadanoff renormalization  group  approximation. The  model
with  $p=3$  and  $M=3$  has  at  mean-field  level  the  ideal  glass
transition at the Kauzmann temperature and at lower temperatures still
the Gardner transition to a state  like that of an Ising spin glass in
a  field.  The  model  with  $p=3$  and $M=2$  has  only  the  Gardner
transition.   In the dimensions  studied, $d=2,3$  and 4,  both models
behave  almost  \textit{identically},   indicating  that  the  growing
correlation length  as the temperature  is reduced in these  models --
the analogue  of the point-to-set  length scale --  is not due  to the
mechanism postulated  in the random  first order transition  theory of
glasses, but is  more like that expected on the  analogy of glasses to
the Ising spin glass in a field.
\end{abstract}



\maketitle

One of  the leading contenders for  a theory of glasses  is the random
first-order  transition theory  (RFOT) \cite{KTW,LW,BB}.   It  had its
genesis in  $p$-spin glass models \cite{KTW}.  The particular $p$-spin
models which might be relevant to the properties of structural glasses
have a mean-field  limit in which there are  two critical temperatures
$T_d$ and $T_K$.  The upper temperature $T_d$ marks the temperature at
which  dynamical singularities  appear  and are  like  those found  in
mode-coupling theory  \cite{MCT}.  The lower temperature  $T_K$ is the
temperature  of the  ideal  glass transition.  This  occurs where  the
configurational entropy  (or log of  the number of  metastable states)
vanishes    \cite{Kauzmann}.    Mean-field   like    calculations   on
glass-forming liquid models support this picture \cite{MP, PZ}.

It has always  been recognized that the dynamical  transition at $T_d$
will disappear outside the mean-field limit due to activated processes
out of the metastable states.  These activated processes make even the
existence of metastable states  problematical. In a recent paper Franz
et al.  \cite{Franz} calculated a dynamical correlation length using a
field theoretic  approach and found  on comparing with  numerical data
that the  agreement was  only good  when the length  scale was  of the
order  of a  particle diameter.  The field  theory predicts  that this
length  scale  should diverge  but  the  simulations  reveal that  the
correlation length  remains small \cite{Kob}, even  though time scales
increase rapidly: the dynamical transition is an example of an avoided
transition.

The ideal glass  transition at $T_K$ at mean-field  level (or infinite
dimension) is  associated with a static (equilibrium)  transition to a
state with one-step replica  symmetry breaking (1RSB) \cite{KTW}.  The
order parameter $q$ jumps from zero in the high temperature phase to a
finite value at and below $T_K$. It is this jump in $q$ which leads to
the \lq  \lq first-order" part  in the name  of the RFOT  theory.  The
configurational  entropy (or complexity)  vanishes as  the temperature
approaches $T_K$ from above.  While there is widespread agreement that
the transition at $T_d$ becomes just a crossover or avoided transition
in finite  dimensions, there is no  consensus about what  happens to the
ideal glass  transition outside the  mean-field limit.  One of  us has
argued that  the 1RSB  transition must also  be avoided in  any finite
dimension \cite{Moore06}, just like the dynamical transition at $T_d$.
 In other words, the lower critical dimension
$d$ of the 1RSB state is infinite.

In this paper we examine  $M$-$p$-spin models; in particular the cases
of $p=3$ with  $M=2$ and $M=3$.  These variants  of the $p$-spin model
have been  extensively studied \cite{YM, DBM, PPR,  Larson, MD}. Their
significance  is that calculations  and simulations  can be  done with
them both  at the mean-field level  and in finite  dimensions.  In the
$M$-$p$-spin model,  there are $M$  Ising spins $\sigma^{(\alpha)}_i$,
$\alpha=1,2,\cdots,M$  on   each  site  $i$  of   (say)  a  hypercubic
lattice.  The   spins  interact  with   each  other  via   a  $p$-body
interaction.   The Hamiltonian  involves terms  of products  of $p$
spins chosen from the spins  in a pair of nearest-neighbor sites.  For
the $p=3$ case, the Hamiltonian is given by
\begin{eqnarray}
\mathcal{H}&=&-\sum_{\langle ij\rangle}\sum^M_{\alpha<\beta}\sum^M_{\gamma}\Big( 
J^{(\alpha\beta)\gamma}_{ij} \sigma^{(\alpha)}_i \sigma^{(\beta)}_i 
\sigma^{(\gamma)}_j \nonumber\\
&&\quad\quad\quad\quad\quad\quad +
J^{\gamma(\alpha\beta)}_{ij} \sigma^{(\gamma)}_i \sigma^{(\alpha)}_j 
\sigma^{(\beta)}_j\Big),
\end{eqnarray}
where the notation $\langle ij\rangle$  means that the sum is over all
nearest neighbor pairs $i$ and  $j$.  The number of different coupling
constants,              $J^{(\alpha\beta)\gamma}_{ij}$             and
$J^{\gamma(\alpha\beta)}_{ij}$ for  given $\langle ij\rangle$  is just
$2M\binom{M}{2}=M^2(M-1)$.   All these  couplings  are usually  chosen
independently from  a Gaussian distribution  with zero mean  and width
$J$. 

The  versions with $M = 2$ and $M=3$ are  of particular  interest  as  at
mean-field  level they have  quite different  kinds of  behavior.  The
model with $M = 3$ at mean-field level has both a dynamical transition
at $T_d$ and an ideal glass  transition at $T_K$.  The model with $M =
2$ is  completely different  at mean-field level.   It has  neither of
these transitions. The origin of the differences can be glimpsed by putting 
the $M$-$p$-spin model  into a field theoretical framework.

The standard way of doing this  is to use the Hubbard-Stratonovich transformation 
on the replicated partition function and then trace over the spins. 
The resulting field theory associated with this model is the following
Ginzburg-Landau-Wilson Hamiltonian 
\begin{eqnarray}
\mathcal{H}_{\mathrm{GLW}}&=&\int d^d\mathbf{r} \Bigg\{
\frac{1}{2}\sum_{a<b}[\nabla q_{ab}(\mathbf{r})]^2+\frac{t}{2}
\sum_{a<b}q^2_{ab}(\mathbf{r}) \nonumber \\
&&\quad\quad-\frac{w_1}{6}\mathrm{Tr} q^3(\mathbf{r}) 
-\frac{w_2}{3}\sum_{a<b} q^3_{ab}(\mathbf{r}) \Bigg\} ,
\label{GLW}
\end{eqnarray}
where $q_{ab}(\mathbf{r})$ is the order  parameter and $a$ and $b$ are
replica indices  running from 1 to  $n$ with $n\to 0$.   At mean field
level,  this  model has  been  known  for a  long  time  to show  very
different behavior depending on the value of $R=w_2/w_1$ \cite{Gross}.
When  $R>1$, there  are two  transitions at  the mean  field  level as
described above; a dynamical  transition at some temperature $T_d$ and
a thermodynamic  transition at  a lower temperature  $T_K$ to  a state
with one-step  replica symmetry breaking. When $R < 1$ neither of these transitions
will occur.  In Ref.~\cite{mpspin}, the  ratio $R$  was evaluated  for the
$M$-$p$-spin model for general values of $M$ and $p$. The cases we are
interested in this paper, namely $p=3, M=2$ and $p=3, M=3$, correspond
to $R\approx 0.879$ and $R=2$, respectively.  Therefore
the two models should indeed show very different mean-field behavior.

 At temperatures below $T_K$ for the  case where $R > 1$, there is yet
 another  transition, the  Gardner transition,  to a  state  with full
 replica symmetry  breaking (RSB) \cite{Gardner}.  For  the case where
 $R<1$, there is at mean-field level only one transition - the Gardner
 transition, a continuous  transition to a state which  is expected to
 have full RSB, (although this has never been checked explicitly to the best of our knowledge).

The transition discovered  by Gardner is thus present  for both the $M
=2$ and  $M=3$ models at mean-field level  \cite{Gardner}.  She showed
that the state with full replica symmetry breaking was very similar to
that of the Ising spin glass  in an applied field $h$. For this model,
there is  a line in the  $h-T$ phase diagram,  the de Almeida-Thouless
(AT)  line,  \cite{AT},   which  separates  the  paramagnetic  replica
symmetric   state  from   the   state  with   full  replica   symmetry
breaking.  Arguments have been  presented \cite{AT6,Moore12}  that the
lower  critical dimension  of   states  with  full replica  symmetry
breaking is 6.  The Gardner transition, 
which is in the same universality class as the AT transition, should  be another
avoided transition for all $d \le 6$.

In this  paper we have studied both  the models with $M=2$  and $M =3$
within the Migdal-Kadanoff (MK) renormalization group approximation in
dimensions  $d=2, 3$  and $4$  to determine  how  thermal fluctuations
modify   the  mean-field  picture   of  these   two  models.   The  MK
approximation is one  of the few approximations which  is reliable for
the  study of  spin  glasses  in low  dimensions  \cite{DBM, YM}.  The
details  of  our  calculation  are  as  in  \cite{DBM,  YM}.   We  are
interested   in  particular   whether  in   the   physically  relevant
dimensions, $d=2$  and $d=3$, whether  there are any vestiges  left of
the  mean-field transitions.  One can  see in  the  molecular dynamics
study  of Kob  et  al.  \cite{Kob}, clear  remnants  of the  dynamical
transition. Only  equilibrium properties are studied in  this paper so
only the remnants of transitions which  could be seen are those of the
ideal glass transition and the Gardner transition for the case with $M
= 3$, and just the Gardner transition for the case $M =2$.

We determined the correlation length  $\xi$ by the same method which was used in Refs. \cite{MD,DBM,YM}, that is via the decay of the interactions  $J_{ij}$ with distance $L$ on the MK hierarchical lattice:
 $J_{ij} \sim \exp(-L/\xi)$.
 As the temperature $T$ is reduced to zero this correlation
length grows to a  value $\xi(0)$, which is strikingly large  especially for $d =
4$. The data on $\xi(0)$ are presented in Table \ref{xi(0)}. The large values of
$\xi(0)$ certainly suggests there is an avoided transition mechanism at work. Even in
$d=3$, $\xi(0)$ is considerably larger than those which have been obtained in simulations of realistic glass models, at least down to temperatures which are currently practical \cite{Kob}.

\begin{table}
\caption{\label{xi0} The correlation length $\xi(0)$ in the zero-temperature limit 
for various values of $M$ and $d$ measured  in units of the lattice spacing.}
\begin{ruledtabular}
\begin{tabular}{cccc}
~ & $d=2$ & $d=3$ & $d=4$ \\
$M=2$ & 11 & 26 & 72 \\
$M=3$ & 12 & 38 & 165
\end{tabular}
\end{ruledtabular}
\label{xi(0)}
\end{table}

\begin{figure}
\includegraphics[width=0.48\textwidth]{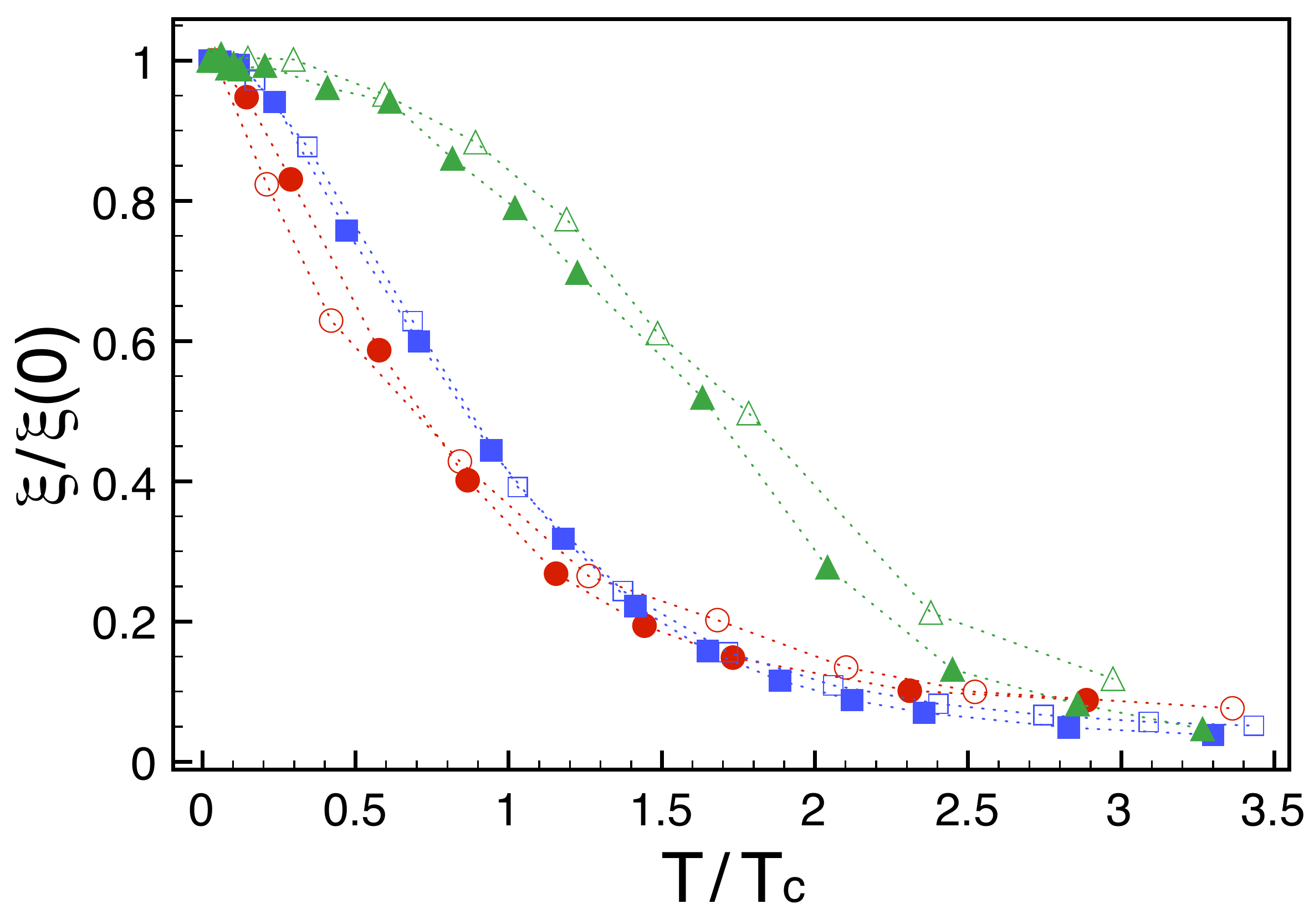}%
\caption{(Color online) Behavior  of  the  scaled correlation  function  $\xi/\xi(0)$
  plotted  versus the  scaled  temperature $T/T_c$.  The closed  symbols
  correspond to $M=3$, the open symbols to $M =2$. The upper (green triangles)
  curves are  for $d=4$,  the middle  curves (blue squares) are  for $d  = 3$,
  while the lower two curves (red circles)  are for $d=2$.}
\label{three_spin}
\end{figure}

It is useful to measure temperature $T$ on the scale of the mean-field
transition temperature  $T_c$ (defined as when the  coefficient $t$ in
Eq.  (\ref{GLW}) equals  zero).  For $M =2$, $T_c  = (\sqrt 2 z)^{1/2}
J$, while for $M=3$, $T_c=(3 z)^{1/2}  J$, where $z =2d$ is the number
of  nearest neighbors  on  the hypercubic  lattice \cite{mpspin}.   In
Fig.~\ref{three_spin} the ratio  of $\xi/\xi(0)$ has been plotted as
a function  of $T/T_c$.  It shows  that as a function  of $T/T_c$, the
ratio $\xi/\xi(0)$ is  essentially the same for both  $M=2$ and $M=3$.
We had expected to see for the  case of $M =3$ features which could be
associated with  a possibly avoided  ideal glass transition  at $T_K$.
None is  visible in Fig.~\ref{three_spin}. This result is  our main
finding. What is its significance?

The  correlation  length  studied   here  is  the  equivalent  of  the
point-to-set length scale  \cite{Cavagna,Biroli,Kob} studied in glassy
supercooled liquids. In the RFOT theory \cite{BB,LW} this length scale
grows  as  the  temperature  is  reduced and  eventually  diverges  at
$T_K$. In that theory the growth of the  correlation length is driven
by the decrease of the configurational entropy (or complexity) to zero
as  the temperature approaches  $T_K$. Since  for $M  =2$ there  is no
ideal glass transition yet this model  is almost identical in
its properties with  those of the $M=3$ model,  the growing correlation
length  cannot be a  remnant of  the ideal  glass transition.  It must
instead be a remnant of the Gardner transition which is
 common to both models.

Thus  the  mechanism behind  the  growing  correlation  length as  the
temperature  is reduced  cannot be  that  envisaged in  the RFOT,  but
instead must  be that associated  with the growing  correlation length
which arises in the Ising spin  glass in a field as the temperature is
lowered.   According   to  the  droplet   picture,  \cite{Fisher-Huse,
  Bray-Moore,  McMillan}  the  correlation  length  increases  as  the
temperature  is decreased and  saturates at  $T=0$ to  a value  set by
equating the interface energy between  a droplet of size $\xi(0)$ and its time reverse,
$\sim J \xi(0)^{\theta}$, to  the energy gained from the  field on flipping
the droplet,  $\sim h \xi(0)^{d/2}$.   The exponent $\theta  \approx -0.28$
for   $d=2$  and   $\theta   \approx  0.24$   for   $d=3$  (see   Ref.
\cite{Boettcher}  for a  review of  the value  of $\theta$  in various
dimensions  $d$). Table  1  shows  that $\xi(0)$  gets  larger as  the
dimensionality goes  up and when there is  an AT line, i.e.  when $d >
6$, it would  be expected to be infinite.  The MK approximation itself
is a low-dimensional approximation (it is exact in one dimension), and
cannot be  trusted to be  even qualitatively correct in  dimensions as
high as six.

We  have argued before  that the  growing (point-to-set)  length scale
seen in supercooled  liquids as the temperature is  reduced \cite{ MY,
  TM} is a consequence of their being in the same \lq \lq universality
class'' as  the Ising spin  glass in a  field. However, until  now, we
could not rule out the possibility that the growing length scale might
arise  through a  1RSB transition  as in  RFOT, (but  possibly avoided
because  of the  mechanism  in Ref.   \cite{Moore06}). The  similarity
between the $M=2$ and  $M=3$ models shown in Fig.~\ref{three_spin} now
removes  that possibility  for  the  $M=3$ model.  In  a recent  paper
\cite{G4}, it  has been argued that the  discontinuous 1RSB transition
in the $M  = 3$ model might be removed  by the fluctuations on short
length scales.


\begin{figure}
\includegraphics[width=0.49\textwidth]{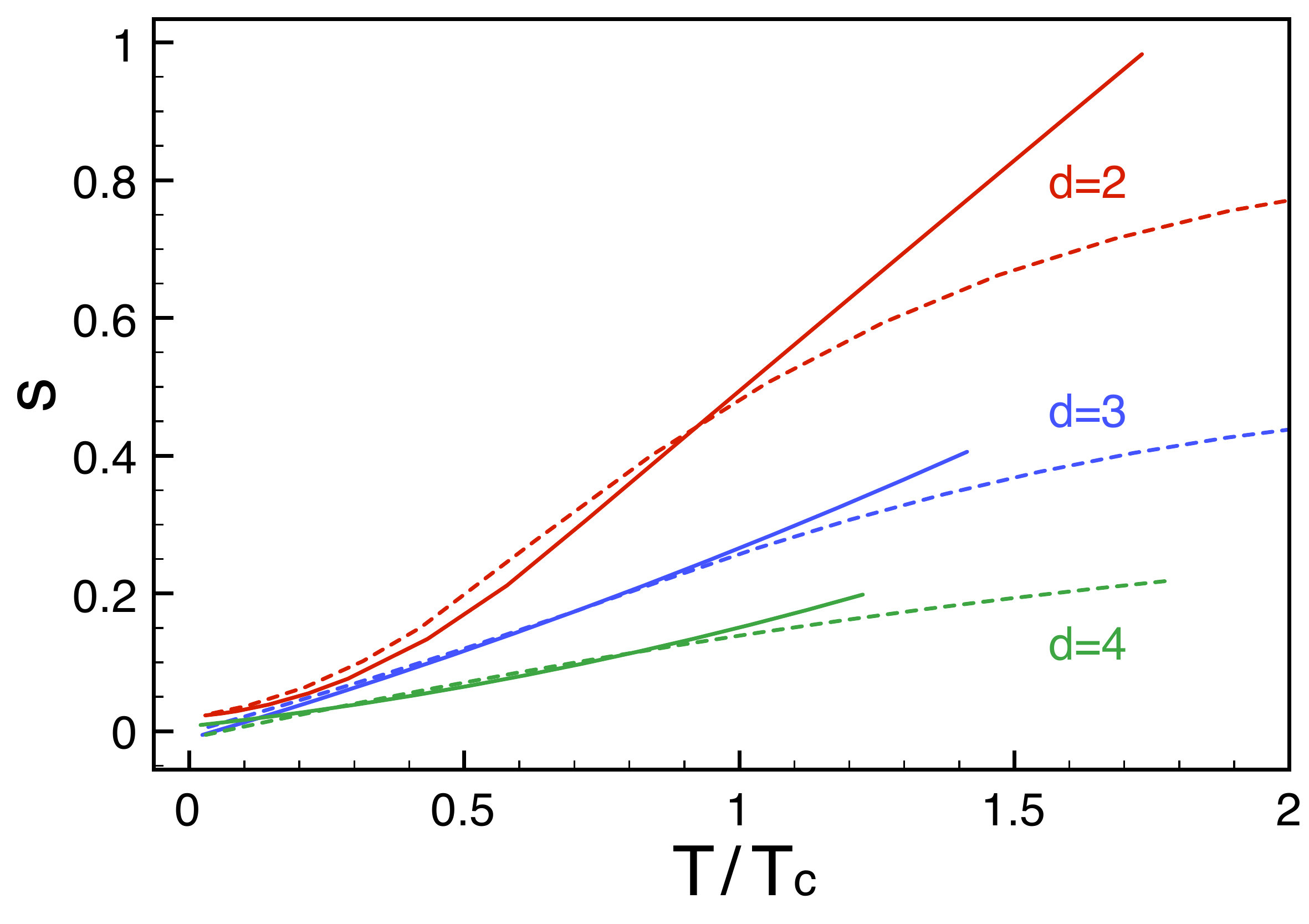}%
\caption{(Color online) The entropy $S$ per site as a function of the scaled temperature $T/T_c$. The solid lines are for the $M=3$ model, the dashed lines are for the $M=2$ model. }
\label{entropy}
\end{figure}

In RFOT theory, the configurational entropy is argued to go to zero at
the Kauzmann temperature $T_K$. For  $M$-$p$ spin glass models it is
not clear how the  configurational entropy should be determined outside
the  mean-field limit,  but we  have  studied their  total entropy  by
numerically differentiating  the free energy calculated  within the MK
approximation  (which leads  to  some inaccuracy  near  $T=0$). It  is
plotted  in  Fig.~\ref{entropy}   as  a  function  of  the  reduced
temperature  $T/T_c$. Once again  the models  with $M  =2$ and  $M =3$
behave almost identically, indicating that when the correlation length
gets  large,  there  is  present  a  form  of  universality.  At  high
temperatures where the  correlation length is small, the  two types of
model have very different entropies: the high temperature limit of the
entropy  per site  is  $k_B  M \ln  2$.   While there  is  no sign  in
Fig. ~\ref{entropy} of  the entropy  vanishing below  some temperature
$T_K$,   the  entropy  is  smaller  at  lower
temperatures for  the $d  =4$ versions of  the model, and  behavior in
four  dimensions is  going  to  be closer  to  mean-field theory  than
behavior in two dimensions.

\begin{figure}
\includegraphics[width=0.49\textwidth]{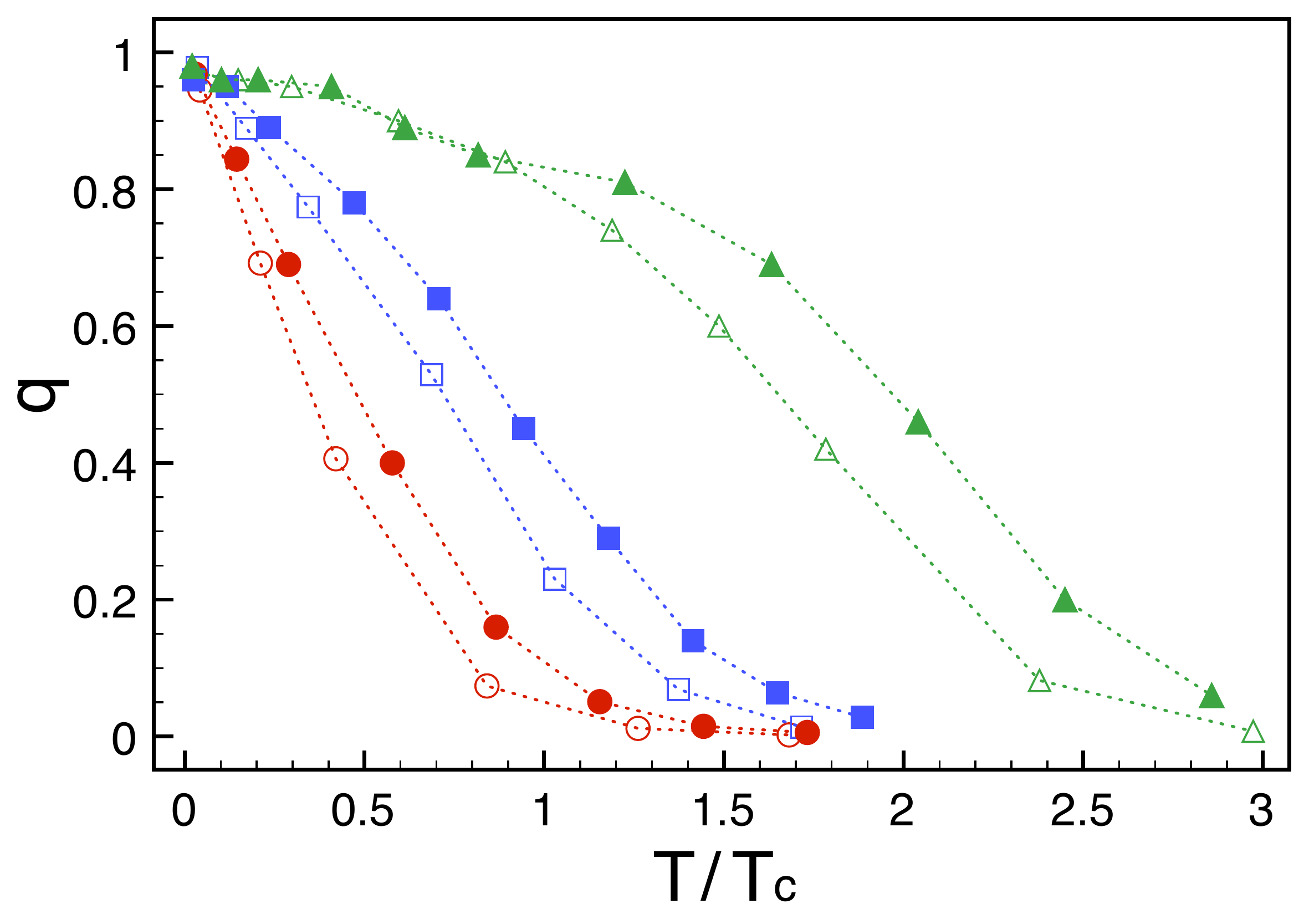}%
\caption{(Color online) The Edwards-Anderson order parameter  $q$ as a function of the reduced temperature.
 The symbols have the same association with $M$ and $d$ as in Fig. \ref{three_spin}.   }
\label{qEA}
\end{figure}

We have also determined the Edwards-Anderson order parameter $q$ 
\begin{equation}
q = [\langle \sigma _i^{(\alpha)} \rangle ^2]_{\rm av},
\end{equation}
where the average  is over the bond realisations.   $q$ is independent
of the value  of $\alpha$, (which runs from  $\alpha=1$ to $M$). Under
the  MK iteration scheme  the couplings  flow to  the high-temperature
fixed point where  the block spins are decoupled  and only single site
terms remain. As a consequence it is easy to evaluate $q$.

The Edwards-Anderson order  parameter $q$ is plotted as  a function of
the reduced temperature  in Fig. \ref{qEA}.  The figure  shows that $q
\to 1$ as $T \to 0$ which is to be expected. At low temperatures where
the correlation  length is large  both the $M  = 2$ and $M  =3$ models
behave almost  identically, which  is another example  of the  \lq \lq
universality" emerging in the problem.

What is striking is that  $q$ is non-zero at any temperature, although
it does become  very small when $T  \gg T_c$. $q$ is a  measure of the
extent  to  which   the  spin  at  site  $i$   remembers  its  initial
orientation: i.e.
\begin{equation}
q =\frac{1}{N}\sum_i\langle \sigma_i^{(\alpha)}(0) \sigma_i^{(\alpha)}(t) \rangle, \hspace{0.5cm} t \to \infty.
\end{equation}
Thus in  the $p$-spin models  the spins never completely  forget their
initial  orientations,  no  matter  how  high  the  temperature.  This
behavior is not  a consequence of using the MK  approximation. It is a
feature    which   arises    in   any    model   described    by   the
$\mathcal{H}_{\mathrm{GLW}}$ of Eq.~ (\ref{GLW}) with a non-zero value
of  $w_2$,  the  term   which  breaks  the  time-reversal  symmetry.  At
mean-field level $q$ does vanish at temperatures above $T_d$.

 $p$-spin models are  meant to be useful models  for understanding the
properties of  supercooled liquids so this feature  of a non-vanishing
$q$  is   hard  to  reconcile  with  the   properties  of  supercooled
liquids. These  forget their  initial conditions after  a time  of the
order of the  alpha relaxation time, so for them $q$  is zero on long
time  scales. Maybe  $p$-spin  models are  useful  for describing  the
properties  of supercooled liquids but only on  timescales less  than the  alpha relaxation
time. Given  the huge effort  which has gone into  investigating their
properties, this is certainly to be hoped.
 
Finally we  note that  if we  had used the  MK approximation  with the
further approximations which were  made in Ref. \cite{CBTT}, a Kauzmann
transition would have been found for the model with $M =3$. It is only
by  carrying out  the MK  calculation  exactly that  one recovers  the
correct behavior \cite{YM}.

JY was supported by WCU program through the KOSEF funded by the MEST
(Grant No. R31-2008-000-10057-0).




\end{document}